\documentclass[twocolumn,showpacs,preprintnumbers,amsmah,amssymb]{revtex4}
\usepackage{graphicx}
\usepackage{dcolumn}
\usepackage{bm}
 \usepackage[latin1]{inputenc}

\begin{document}                

\title{Origin of the reduced exchange bias in epitaxial FeNi(111)/CoO(111) bilayer}

\author{F. Radu$^{1, \footnote{Florin.Radu@helmholtz-berlin.de}}$, S. K. Mishra$^1$, I. Zizak$^1$, A. I. Erko$^1$,  H. A. D\"urr$^1$ and W. Eberhardt$^1$}
\affiliation{$^{1}$ Helmholtz-Zentrum Berlin f\"ur Materialien und
Energie, Albert-Einstein Strasse 15, D-12489, Berlin, Germany}
\author{G. Nowak$^{2}$, S. Buschhorn$^{2}$,  K. Zhernenkov$^{2,3}$, M. Wolff$^{2,3}$,  and H. Zabel$^2$}
\affiliation{$^{2}$Institut f\"{u}r
Experimentalphysik/Festk\"{o}rperphysik, Ruhr-Universit\"{a}t
Bochum, D-44780 Bochum, Germany} \affiliation{$^{ 3}$Institut
Laue-Langevin, F-38042 Grenoble Cedex 9, France}
\author{D. Schmitz$^{4}$,  E. Schierle$^{4}$,  E. Dudzik$^{4}$, and R. Feyerherm$^{4}$}
\affiliation{$^{4}$Helmholtz-Zentrum Berlin f\"ur Materialien und
Energie, Glienicker Str. 100, D-14109 Berlin, Germany}
\date{\today}

\begin{abstract}
We have employed Soft and Hard X-ray Resonant Magnetic Scattering
 and Polarised Neutron Diffraction  to study the magnetic
interface and the bulk antiferromagnetic  domain state of the
archetypal epitaxial Ni$_{81}$Fe$_{19}$(111)/CoO(111) exchange
biased bilayer. The combination of these scattering tools provides
unprecedented detailed insights into the still incomplete
understanding of some key manifestations of the exchange bias
effect. We show that the several orders of magnitude difference
between the expected and measured value of exchange bias field is
caused by an almost anisotropic in-plane orientation of
antiferromagnetic domains. Irreversible changes of their
configuration lead to a training effect. This is directly seen as
a change in the magnetic half order Bragg peaks after
magnetization reversal. A 30~nm  size of antiferromagnetic domains
is extracted from the  width  the (1/2 1/2 1/2) antiferromagnetic
magnetic peak measured both by neutron and x-ray scattering. A
reduced blocking temperature
 as compared to  the measured antiferromagnetic ordering temperature
clearly corresponds to the blocking of  antiferromagnetic domains.
Moreover, an excellent correlation between the size of the
antiferromagnetic domains, exchange bias field and frozen-in spin
ratio is found, providing a comprehensive understanding of the
origin of exchange bias in epitaxial systems.

\end{abstract}
 \maketitle

\section{Introduction}

 Although the first exchange biased
system was engineered by nature a few billion years
ago~\cite{mcenroe:2007}, its observation has been possible only 60
years ago by Meiklejohn and Bean~\cite{bean:1956}, when studying
Co particles embedded in their natural oxide (CoO) matrix. After
the discovery of the Giant Magnetoresistance (GMR)
effect~\cite{baibich:1988,binasch:1989} the exchange bias (EB)
effect has become an integral part of modern magnetism with
implications for basic research and for numerous device
applications like magneto-electronic switching devices
(spin-valves) and for random access magnetic storage units. For
these applications a predictable, robust, and tunable exchange
bias effect is required.

The EB effect manifests itself in a shift of the hysteresis loop
in the  negative or the positive direction with respect to the
applied field. Its origin is related to the magnetic coupling
across the common interface shared by a ferromagnetic (FM) and an
antiferromagnetic (AF) layer. Extensive research  is being carried
out to unveil the microscopic origin of this effect
\cite{berkowitz:1999,nogues:1999,stamps:2000,kiwi:2001,rz:2008,iglesias:2008}.
The details of the EB effect depend crucially on the AF and on the
interface separating it from the FM layer. However, some
characteristic features are still under debate: (1) The size of
the exchange field is up to several orders of magnitude lower than
expected for many epitaxial systems with an uncompensated AF
surface; (2) exchange bias field ($H_{EB}$) and coercive field
($H_c$) increase as the system is cooled in an applied magnetic
field below the blocking temperature ($T_B$) of the AF layer with
$T_B \le T_N$, where $T_N$ is the N\'{e}el temperature of the AF
layer; (3) the magnetization reversal can be different for the
ascending and descending part of the hysteresis
loop~\cite{nikitenko:2000,fitzsimmons:2000,radu:2002:1,lee:2002,gierlings:2002,brems:2005};
(4)$H_{EB}$  and $H_c$ can vary when hysteresis loops are measured
consecutively, a phenomenon called training
effect~\cite{paccard:1966}. Furthermore, a positive $H_{EB}$ has
been observed after cooling an AF/FM system in very high magnetic
fields at low temperatures~\cite{nogues:1996,hong:1998} and close
to the blocking
temperature~\cite{radu:2003:1,gredig:2002,ali:2007}. More than 27
theoretical models have been developed for describing possible
mechanisms of the EB effect. The main motivation for most of them
is to describe the discrepancy between the measured versus
predicted value for the $H_{EB}$ and $H_c$.

We address this discrepancy  by studying  an epitaxial
Ni$_{81}$Fe$_{19}$(111)/CoO(111) exchange biased bilayer by
polarized neutron and x-ray scattering and reflectivity. We show
that the exchange bias for an epitaxial Ni$_{81}$Fe$_{19}$/CoO is
several orders of magnitude less that expected due to the
particular domain state of the AF layer. The available coupling
energy is transformed in coercivity, mediated by the  magnetically
disordered interface. The blocking temperature of the exchange
bias appears as the blocking of the AF domains, as revealed  by
neutron scattering. The temperature behavior of the frozen-in and
rotatable AF spins are compared to the EB field and average domain
sizes.

The paper is organized as follows: in Sec. II we describe the
sample growth and show the structural characterization by
 x-ray diffraction utilizing  synchrotron radiation. In Sec. III we
show the formation of AF domains  by analyzing the (111) and (1/2
1/2 1/2) charge and magnetic peaks, respectively. For these
measurements we have used
 Resonant X-ray Diffraction  at the Co K-edge.
In Sec. IV. we study the magnetization reversal of the
ferromagnetic layer by Polarised Neutron Reflectivity. Using the
same geometry, we further characterize the average orientation of
the antiferromagnetic domains by Polarised Neutron Diffraction.
Moreover the temperature dependence of the averaged AF domain size
is extracted from the transverse (1/2 1/2 1/2) magnetic Bragg
peak. In Sec. V. we show the temperature dependence of the
uncompensated spins  measured by Soft X-ray magnetic Scattering at
the Co L$_3$ edge. In the same section we discuss the correlation
between the AF domain sizes, uncompensated spins and exchange bias
as a function of temperature. In Sec. VI we provide the
conclusions of our study.

\section{Sample growth and characterization}

The samples have been grown by dc-magnetron sputtering (at BESSY)
in an Argon atmosphere of~$1.5\, x \, 10^{-3}$ mbar with a base
pressure of $2\, x \, 10^{-8}$ mbar. Unlike the previously grown
CoO layers, where  rf-sputtering was preferred due to the
insulating nature of the CoO target,  dc-magnetron sputtering
offers the advantage of higher ´deposition rates and, therefore a
thicker CoO layer can be grown in stable conditions. Five
substrate crystals have been used to test the quality of the CoO
films: MgO(100), MgO(110), MgO(111), Al$_2$O$_3$(0001), and
Al$_2$O$_3$(11$\bar{2}$0). Although the last three substrates
provide the (111) uncompensated surface for CoO, the highest
structural quality was  achieved by using an
Al$_2$O$_3$(11$\bar{2}$0) crystal. The substrate was rinsed in
ethanol and cleaned in an ultrasonic bath for 30 minutes. After
annealing to 700~$^\circ$C for 15 minutes, the temperature was
decreased to 500~$^\circ$C where a 2000~\AA~thick CoO layer was
grown. Afterwards, the temperature was further decreased to room
temperature for the deposition of  a 120~{\AA} Ni$_{81}$Fe$_{19}$
(Permalloy$\equiv$Py ) film. To prevent oxidation, a 50~{\AA} Au
capping layer was grown on top of the bilayer. The reduced
deposition temperature for the Py and Au layers was chosen in
order to reduce temperature induced interdiffusion at the
interface.

\newpage
\begin{figure}[ht]
         \includegraphics[clip=true,keepaspectratio=true,width=.8\linewidth]{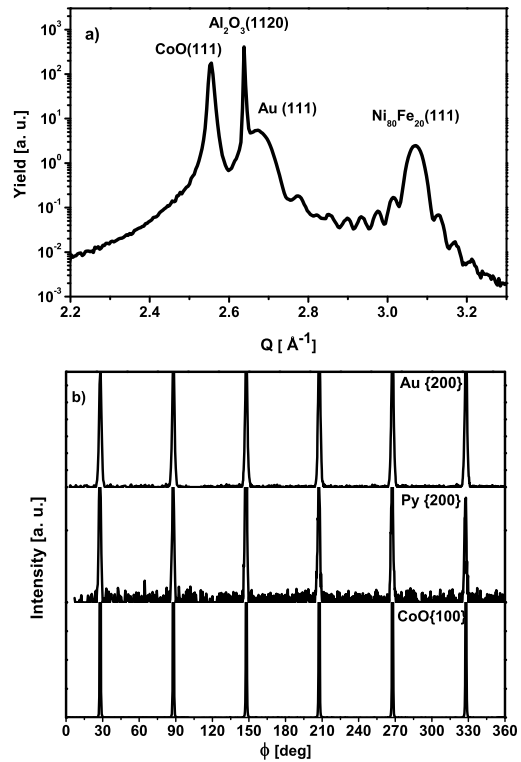}
          \caption{a) Longitudinal  x-ray diffraction along the [111] crystallographic axes. All layers exhibit a Bragg peak. The Py and Au layers show Laue oscillations
           suggesting an exceptionally  smooth CoO/Py interface.
           b) Orientation distribution of  [100] plane normals
           measured by using asymmetric x-ray diffraction.The epitaxial relation between the CoO, Py
           and Au layer is
           CoO[1$\bar{1}$0]$||$Py[1$\bar{1}$0]$||$Au[1$\bar{1}$0]. All diffractograms were recorded at $E=8048$ eV
($\lambda=1.5405$~\AA).}
   \label{fig1}
 \end{figure}

The structural quality of the samples  was studied by using x-ray
scattering at the MAGS~\cite{dudzik:2006} and
KMC2~\cite{erko:2000} x-ray beamlines at BESSY (Fig.\ref{fig1}a)
and Fig.\ref{fig1}b), respectively) using $\lambda$~=~1.5405~\AA.
Preliminary diffraction measurements were done in the BESSY
Crystallography Laboratory at the two-crystal x-ray diffractometer
TRS-1. A longitudinal Bragg scan is shown in Fig.~\ref{fig1}a. The
CoO peak at Q=2.549~{\AA}$^{-1}$ occurs at the tabulated value
suggesting a stoichiometric growth~\cite{nowak:2007}. The Py peak
at Q=3.07~{\AA}$^{-1}$ exhibits Laue oscillations which are
indicative of
 an excellent smoothness of the Py/CoO interface.
Even the Au capping layer peak at Q=2.67~{\AA}$^{-1}$ exhibits two
Laue oscillations, one at Q=2.77~{\AA}$^{-1}$ and another one at
Q=2.57~{\AA}$^{-1}$, below the CoO peak. In order to probe the
epitaxy relations between the layers we have measured azimuthal
scans around the [100] crystallographic orientation, which makes
an angle of 35.26$^\circ$ with respect to the sample surface. Six
fold symmetry  indicates that the CoO layers consist of at least
two crystallographic domains~\cite{gokemeijer:2001} rotated
60$^\circ$ with respect to each other. The epitaxial  relation
extracted from these data can be expressed as :
CoO[1$\bar{1}$0]$||$Py[1$\bar{1}$0]$||$Au[1$\bar{1}$0] and
corresponds to the Nishiyama-Wassermann epitaxial
growth~\cite{nishiyama:1934,nowak:2007}.

\section{Observation of antiferromagnetic  domains by Resonant X-ray Magnetic Scattering}

The domain formation in the AF CoO layer was studied using x-ray
magnetic scattering at the Co K-edge. Measurements were done at
the 7 T multipole wiggler beamline MAGS, operated by the
Helmholtz-Zentrum Berlin at the synchrotron source BESSY
II~\cite{dudzik:2006}. The sample was cooled from 300 to 10 K in a
800~Oe magnetic field applied parallel to the sample surface. The
peak shape of the structural (111) and the AF ($\frac{1}{2}
\frac{1}{2} \frac{1}{2} $) Bragg peaks was measured both in the
($\theta${/}$2\theta$) and the transverse ($\theta$ rocking scan)
geometries. Linear polarisation analysis with  Au(111) crystal was
used to separate structural and magnetic contributions.

\begin{figure}[ht]
           \includegraphics[clip=true,keepaspectratio=true,width=1\linewidth]{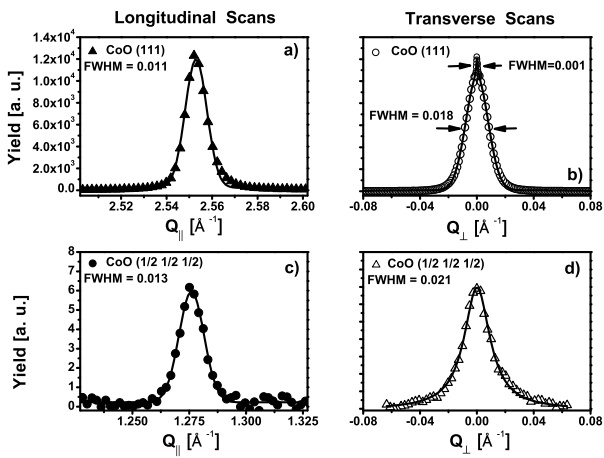}
          \caption{Resonant X-ray Magnetic Scattering at the Co K edge: a) longitudinal and b) transverse (111) Bragg
          peaks; c) longitudinal and d) transverse ($\frac{1}{2} \frac{1}{2}
\frac{1}{2} $) Bragg peaks. The structural (111) peaks provide
information about in-plane and out-of-plane charge correlation.
The longitudinal and transversal half order peaks ar wider as
compared to charge scattering, suggesting the formation of AF
domains. The width of the magnetic transverse scan provides the
in-plane magnetic coherence length for the AF domains. The scans
have been measured at 10~K after  cooling the sample from above
room temperature in an external magnetic field.}
   \label{fig2}
 \end{figure}

Comparing the structural to the magnetic longitudinal peaks
(Fig.~\ref{fig2}a versus Fig.~\ref{fig2}c) we observe an increase
of the full width at half maximum (FWHM) from 0.011~\AA$^{-1}$ to
0.013~\AA$^{-1}$, respectively.
The width of the magnetic peak ($\frac{1}{2} \frac{1}{2}
\frac{1}{2} $) becomes wider as compared to the structural one
(111), suggesting formation of AF domains. Scattering at the
magnetic inhomogeneities provided by the domain walls will
diminish the average coherence length, providing an increased
width of the longitudinal ($\frac{1}{2} \frac{1}{2} \frac{1}{2} $)
magnetic peak with respect to its (111) charge scattering
counterpart.

The transverse scan shown in Fig.~\ref{fig2}d probes
 the in-plane average size of AF domains, often identified as the
magnetic coherence length ($L_{AF}=2 \pi
/FWHM$)~\cite{borchers:1995,bea:2008}. An almost Lorentzian shaped
transverse scan (Fig.~\ref{fig2}d)  is indicative for  a broad
distribution of domain sizes with a mean value of
L$_{AF}\approx$~30~nm. The widths are free of instrumental
resolution. Its shape is also quite different from the structural
(111) transverse scan (Fig.~\ref{fig2}b). A sharp coherent
contribution  is clearly visible on  top of a much broader diffuse
peak. The presence of the sharp peak confirms a high film quality,
but random vacancies and stacking faults contribute to the broad
diffuse charge scattering which becomes more prominent as the film
grows thicker~\cite{csiszarphd:2005}. Notice that the correlation
lengths
   for both structural and
magnetic peaks are very close in magnitude,  suggesting that the
AF domain size is only slightly smaller as compared to the grain
size.

\section{Magnetization reversal and the antiferromagnetic domain state by Polarised Neutron Reflectivity  and Polarised Neutron Diffraction}

Having established the existence of AF domains in the AF layer, we
study now their average in-plane orientation using neutrons.
Polarised neutron reflectivity(not shown) (sensitive to
ferromagnetism) and diffraction (sensitive to the
antiferromagnetism) have been performed at the ADAM diffractometer
at the Institut Laue Langevin, Grenoble~\cite{wolff:2007}. Taking
advantage of the large scattering angles available (2$\theta$:
0-125$^\circ$), we have accessed the half-order Bragg peak
($\frac{1}{2}$ $\frac{1}{2}$ $\frac{1}{2}$) and measured spin
analyzed reflection under the same conditions as low angle neutron
reflectivity. The (111) Bragg peak was not accessible due to the
large neutron wavelength ($\lambda$=4.41~\AA) available for this
experiment.

\subsection{Polarised Neutron Reflectivity study of the
ferromagnetic layer}

\begin{figure}[ht]
          \includegraphics[clip=true,keepaspectratio=true,width=1\linewidth]{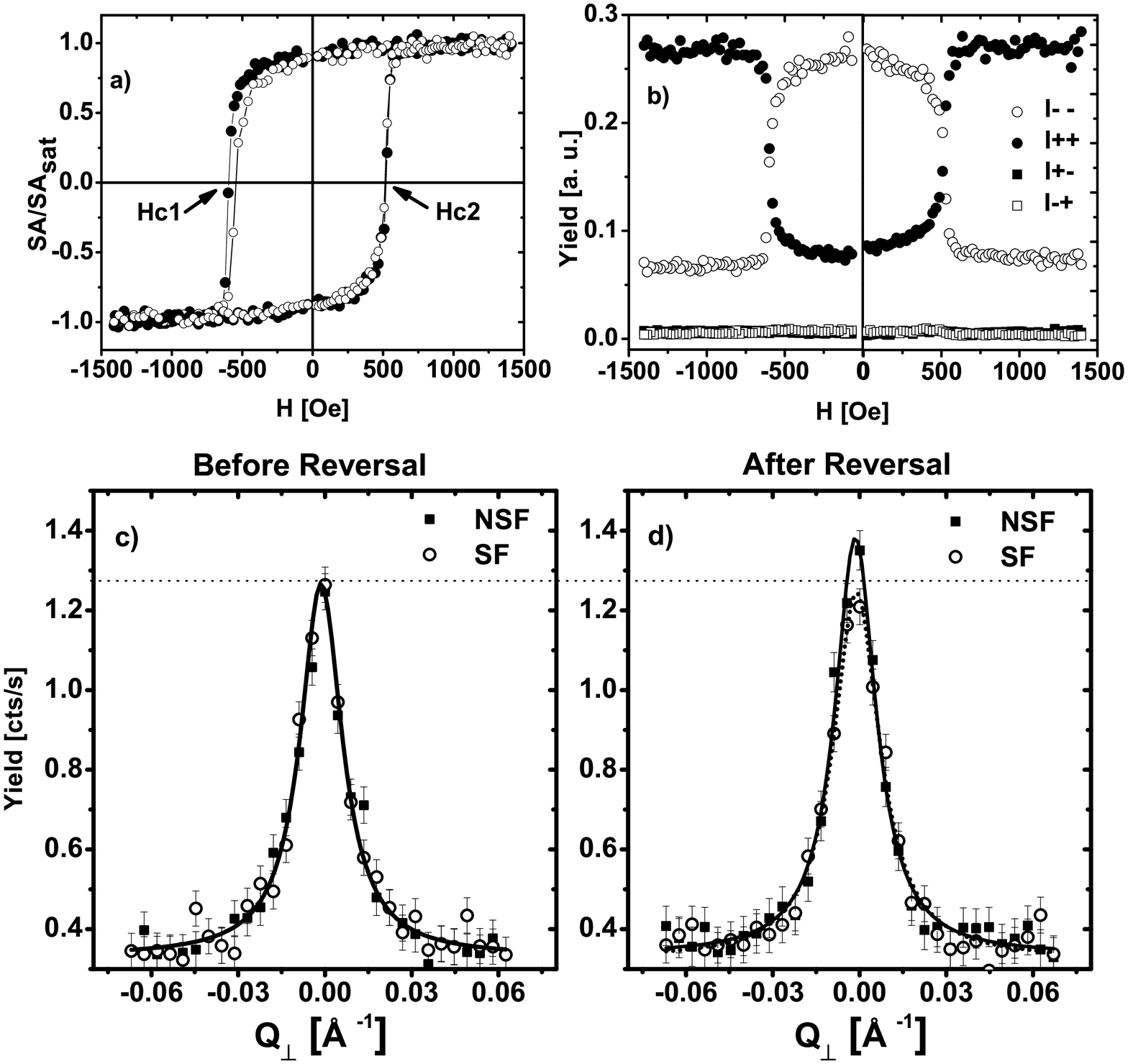}
          \caption{The hysteresis loops a) and the magnetization reversal b) of the ferromagnetic layer. The NSF and SF scattering at the ($\frac{1}{2} \frac{1}{2}
\frac{1}{2}) $ antiferromagnetic half order Bragg peak: c) virgin
state after field cooling from room temperature to 10~K and d) the
trained state after reversing once the magnetization. Both sets of
data (c and d) are measured in an external field of 2~kOe.}
   \label{fig3}
 \end{figure}

In Fig.~\ref{fig3} the magnetization reversal and the hysteresis
loops are shown. The sample has been cooled from above the
N{\'e}el temperature of CoO (T$_N$=291~K) to 10~K in an external
magnetic field of 2~KOe to establish a unidirectional anisotropy.
After  field cooling in saturation, polarized neutron reflectivity
curves were measured at 10~K to find the geometrical conditions
(incident and outgoing angles) for maximum magnetic
contrast~\cite{radu:2003:1} for the ferromagnetic layer. At this
fixed geometry, spin-flip (SF) (I$^{+-}$,I$^{-+}$) and non
spin-flip (NSF) (I$^{++}$,I$^{--}$) reflected intensities were
measured by sweeping the magnetic field. The SF intensities sense
the magnetization component perpendicular to the applied field and
scattering plane, whereas the NSF reflectivities is sensitive to
the magnetization components parallel to the applied field. The
field where the NSF reflected intensities are equal defines the
coercive fields H$_{c1}$ and H$_{c2}$, whereas the SF provides
information on the magnetization reversal. We observe that on both
sides of the hysteresis loop the remagnetization process of the
ferromagnetic layer proceeds by domain wall movements. This is
seen as vanishing SF intensities at the coercive fields. A
rotation of the magnetization would lead to a strong increase of
the SF reflectivity which is absent at both legs of the hysteresis
loop. This contrasts with earlier observations, where an
asymmetric reversal has been observed, albeit for a much thinner
AF layer~\cite{radu:2003:1,brems:2005}. Defining the spin
asymmetry as: $SA(B)=I^{++}-I^{--}/(I^{++} + I^{--} + I^{+-} +
I^{-+})$, the normalized spin asymmetry ($SA(B)/SA(B_{sat})$)
allows us to measure a hysteresis loop.  By measuring a second
consecutive hysteresis loop, we observe that the system exhibits a
small but clear training effect. The characteristics of the
hysteresis loops are: a) the magnetization reversal proceeds via
domain nucleation and propagation for the first and all
consecutive loops; b) the exchange bias field is several orders of
magnitude lower than predicted (H$_{EB}\approx$~20~Oe) and the
coercive field is high H$_{c}\approx$~650~Oe. The exchange bias
field is predicted to be H$_{EB}\approx$~2000~Oe, whereas the
coercive field should not differ from the intrinsic value for the
Py layer which is about 5~Oe~\cite{bean:1956,rz:2008}; c) a small
training effect is clearly seen by comparing two consecutive
hysteresis loops shown in Fig.~\ref{fig3}a. Further hysteresis
loops  exhibits weaker relative changes (not shown).

\subsection{Polarised Neutron Diffraction study of the
antiferromagnetic layer}

Fig.~\ref{fig3}c and Fig.~\ref{fig3}d show polarised neutron
measurements which are sensitive only to the antiferromagnetic
layer, taken after field cooling the sample from above T$_N$ to
10~K, before and after reversing the magnetization. Spin analyzed
transverse scans were measured at the magnetic ($\frac{1}{2}$
$\frac{1}{2}$ $\frac{1}{2}$) reciprocal point (Fig.~\ref{fig3}c).
We mention that a $\lambda/$2 contamination of the neutron beam
can be excluded due to the vanishing intensity of the half order
peak above  T$_N$ (see Fig.~\ref{fig5}a). The transverse scans
carry information about the average antiferromagnetic domain size
and average orientation. The first observation is that after field
cooling the NSF and SF cross-sections are practically equal (see
Fig.~\ref{fig3}c). This translates into almost equally populated
\{111\} domains with a virtually anisotropic in-plane distribution
of the AF spins. On average, an equal number of AF spins are
oriented parallel and perpendicular to the ferromagnetic spins,
respectively.  We have calculated a lateral coherence length of
$\approx$~30~nm, in agrement with the Resonant X-ray Magnetic
Scattering results above (compare to Fig.~\ref{fig2}d). As a
result, the cooling state of the Py/CoO system acquires a
noncollinear magnetic state with a virtually anisotropic in-plane
distribution of the AF domains, while the FM spins are aligned
with the external field.

 \begin{figure}[ht]
          \includegraphics[clip=true,keepaspectratio=true,width=1\linewidth]{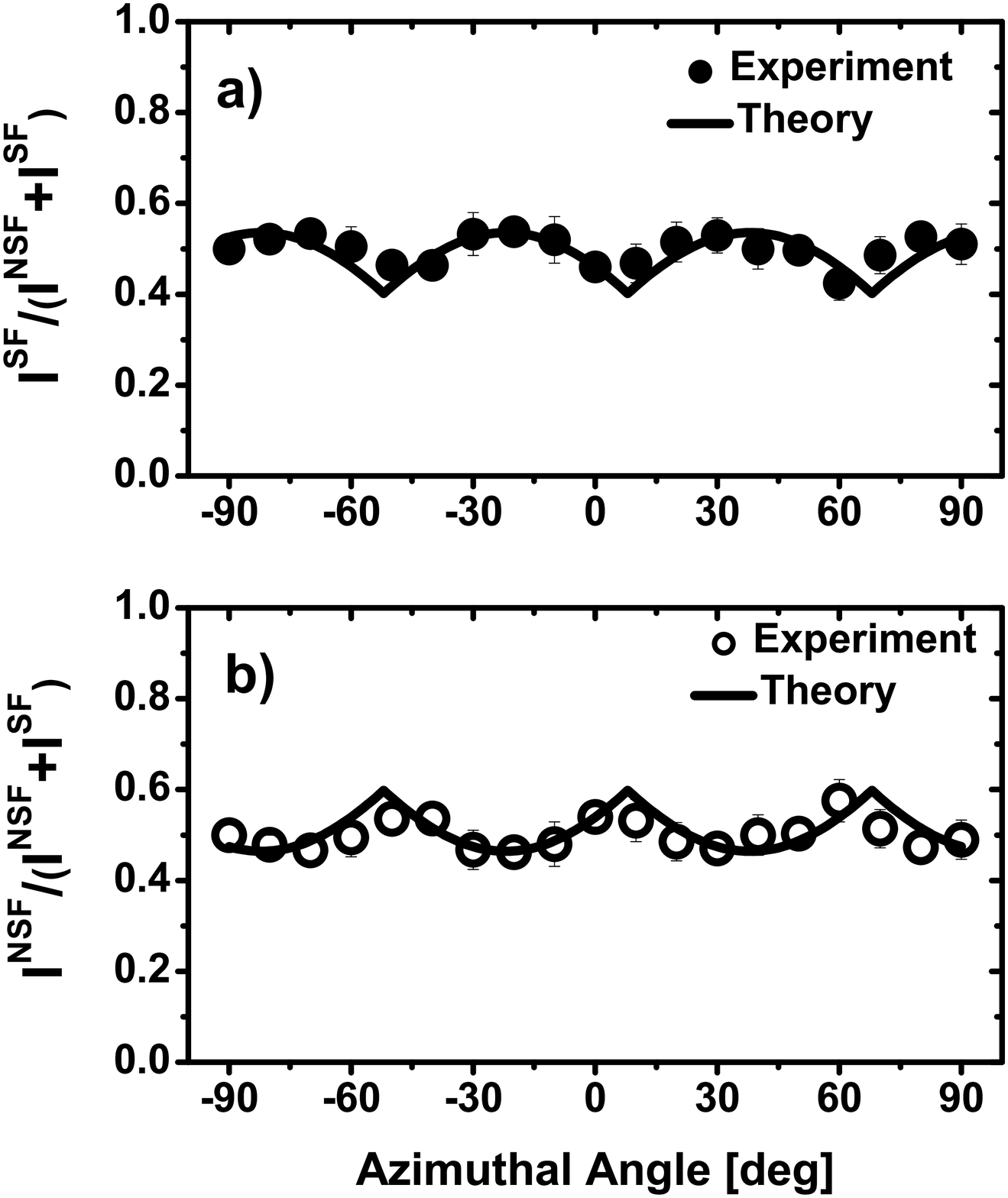}
          \caption{ Normalised a) spin flip
          $I^{SF}/(I^{NSF}+I^{SF})$and  b) non-spin flip
          $I^{NSF}/(I^{NSF}+I^{SF})$ ($\frac{1}{2} \frac{1}{2}
\frac{1}{2} $) Bragg peak integrated intensities measured as a
function of the azimuthal angle. The lines correspond to the
theoretical expectations (calculated Eqs.~\ref{eq1} with n=3 and
$\gamma=\pi/3$)  when assuming a six fold AF spin axes symmetry,
in agreement with the structural data shown in Fig.~\ref{fig1}b.
This graph shows that the AF spins are \textit{not} randomly
oriented in the film plane, but aligned with the crystallographic
axes.}
   \label{fig4}
 \end{figure}

After reversing  the magnetization at H$_{c1}$, the magnetic state
of the AF layer may change as suggested indirectly by the magnetic
interfacial roughness measured before~\cite{radu:2002:1}, leading
to a training effect. By measuring the magnetic Bragg peak of the
AF we access now directly the stability of the AF domain structure
upon reversal. The spin-analyzed transverse scans shown in
Fig.~\ref{fig3}d were measured after a complete hysteresis loop.
The external conditions for the scans before and after reversal
are identical, therefore, we may directly compare the virgin state
of the AF layer after field cooling to the trained one.
Surprisingly, after reversal the AF domain state does undergo
irreversible changes. Under the influence of a strong direct
interfacial coupling some domains appear to rearrange  towards
their stable configuration. The NSF intensity becomes stronger at
the expense of SF scattering. This directly demonstrates that  one
origin of the training effect can be attributed to AF domain
reorientation upon magnetization reversal.

Now we describe an experiment (shown in Fig.~\ref{fig4}) which can
distinguish between an anisotropic in-plane AF spins orientations
and a random AF spin orientational distribution. By an anisotropic
in-plane orientations we understand that the AF spins may  be
directed preferentially parallel to the anisotropy axes provided
by the crystallographic axes, whereas a random orientational
distribution of the AF spins would exhibit no preferential
in-plane orientation. To this end we  performed an azimuthal scan
by measuring SF and NSF neutron integrated intensities at the
($\frac{1}{2}$ $\frac{1}{2}$ $\frac{1}{2}$) Bragg position while
rotating the sample around its normal. Knowing that the scattering
is a coherent process and that the SF probability is sensitive to
the projection of the spin direction onto the SF axis  but not to
the absolute orientational angle, one would expect the normalised
SF and NSF integrated intensities to be equal to:
\begin{eqnarray}
\label{eq1}
\frac{I^{SF}}{I^{SF}+I^{NSF}}=(\sum^{n-1}_{0}{|\sin{(\phi-\gamma*n)}|})^2
\\
\frac{I^{NSF}}{I^{SF}+I^{NSF}}=1-(\sum^{n-1}_{0}{|\sin{(\phi-\gamma*n)}|})^2,\nonumber
\end{eqnarray}
where the integer $n=\pi/\gamma$ is the symmetry number, $\gamma$
is the symmetry angle of the anisotropy axes, and $\phi$ is the
azimuthal angle with $\phi=0$ defining the direction of an AF
anisotropy axis. To obtain the equations above we also used a
conservation law constraining the sum of the NSF and SF
intensities to be constant as a function of the azimuthal angle.
Assuming the AF spins to be oriented along the crystallographic
(anisotropy) directions and making use of the structural data
shown in Fig.~\ref{fig1}b,  we extracted $\gamma=\pi/3$ and $n=3$.
For this case, a SF and NSF integrated intensities modulation
reflecting the six fold structural symmetry  should be observed.
For the other case, of randomly in-plane oriented AF spins, the
normalised SF and NSF yields should show a straight line as a
function of the azimuthal angle. In Fig.~\ref{fig4} the
experimental normalized SF and NSF integrated intensities are
plotted as a function of the azimuthal angle ($\phi$). The sample
has been cooled down to 10~K in an external field of 2~KOe applied
almost parallel to one of the AF anisotropy axis. The field was
reversed as to induce the training effect. Then, the external
field was reduced to about 50~Oe. We observe clear oscillations
with a periodicity of 60 degrees for both NSF and SF signals. The
excellent agreement between the expected values (calculated by
Eqs.~\ref{eq1} with n=3 and $\gamma=\pi/3$) based on the
crystallographic data and the experimental observations in
Fig.~\ref{fig4} leads us to the conclusion that the AF spins
 follow closely the
anisotropy axes. They are not randomly oriented in-plane.

These anisotropic orientations of the AF domains provide on
average a virtually compensated interface and, therefore, the
exchange bias is several orders of magnitude lower than expected.
An AF domain state is  at the core of the Domain State
model~\cite{miltenyi:2000,nowak:2001,nowak:jmmm:2002,nowak:2002,beckmann:2003,misra:2004,scholten:2005,beckmann:2006}.
Their orientation is parallel to the anisotropy axis of the AF
layer which has an unique direction. Here we observe
experimentally a more complex configurations of AF domains, with
orientations  distributed in-plane and parallel to the three
anisotropy axes. The Spin Glass (SG)
model~\cite{rz:2008,radu:jpcm:2006} predicts a reduced AF
anisotropy at the interface. This assumption may help to
understand this particular domain state. Upon field cooling, the
AF acquires  its intrinsic domain state inside the film and by
further cooling this configuration propagates towards the surface,
minimizing the role of interfacial coupling.



To further confirm the influence of the AF domains on the exchange
bias we have measured temperature dependent AF Bragg peaks (not
shown). They provide information on the antiferromagnetic order in
a rather straightforward manner: above the N\'{e}el temperature
the intensity of the half order peak peak vanishes (see
Fig.~\ref{fig5}a), whereas at temperatures for which the long
range AF order is established it acquires non vanishing values.
The integrated peak intensity as a function of temperature
provides the order parameter and the N\'{e}el temperature, which
is  291~K for this sample, in agreement with earlier bulk
measurements~\cite{roth:1958}. Moreover, the width of the AF peak
shown in Fig.~\ref{fig5}b containes additional information on the
temperature dependence of the average AF domain size and the onset
of their stability. The FWHM increases linearly as the temperature
decreases, which translates into a smaller average domain size at
low temperatures. The domain size evolution as function of
temperature may be correlated with an increase of the wall width
(grain boundary). This  may be understood as an interplay between
AF anisotropy and stiffness strengths. For instance, assuming
random fields, Malozemoff
model~\cite{malozemoff:1987,malozemoff:1988:1,malozemoff:1988:2}
provides an analytical dependence for the characteristic length of
the AF domains, namely $L_{AF} \approx \sqrt{(A_{AF}/K_{AF})}$,
where $A_{AF}$ is the exchange stiffness and $K_{AF}$  is the
anisotropy of the antiferromagnet. Assuming that the anisotropy
$K_{AF}$ grows faster as compared to the exchange stiffness
$A_{AF}$, one would expect a shrinking domain size when decreasing
the temperature. The other situation, when $A_{AF}$  grows faster
as compared to $K_{AF}$, would lead  to an increased  AF domain
size. Our experiments show a shrinking average domain size at low
temperatures. This allows us to suggest that an  anisotropy
increase towards lower  temperatures predominantly governs  the
average AF domain size. Another striking behavior is the
observation of a characteristic temperature where the AF domains
reach stability against the exchange interaction with the FM
layer. This blocking temperature is T$_B$=280~K and is lower then
the N\'{e}el temperature. This correlates remarkably well with the
blocking temperature for the exchange bias, to be discussed
further below. By contrast, the N\'{e}el temperature is the
critical temperature which defines the onset of long range spin
order (against thermal fluctuations).

\section{Temperature dependence, frozen-in uncompensated spins, blocking temperature}

\begin{figure}[ht]
          \includegraphics[clip=true,keepaspectratio=true,width=.8\linewidth]{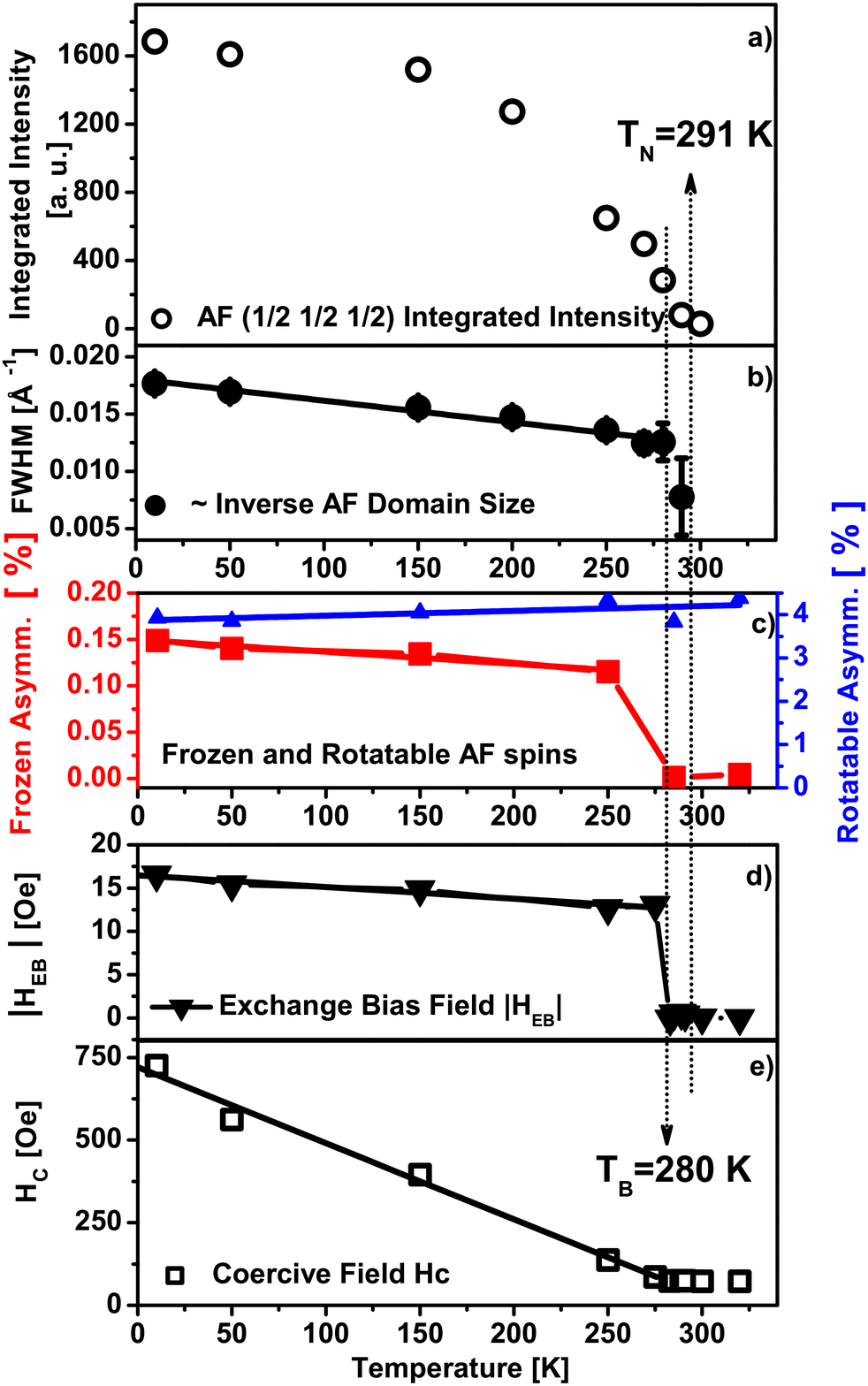}
          \caption{Temperature dependence of
          integrated intensity of the AF magnetic peak (a), in-plane AF coherence length (b), frozen-in and rotatable AF spins
          (c), and exchange bias (e) and coercive (d) fields.}
   \label{fig5}
 \end{figure}

Soft x-ray magnetic scattering measurements were performed at the
UE46 HZB End Station  (Fig.~\ref{fig5}c) and Alice
diffractometer~\cite{grabis:2003} (Fig.~\ref{fig5}d and
Fig.~\ref{fig5}e) operated at BESSY. By tuning the energy close to
the Co L$_3$ absorption edge,   we have measured reflectivity
curves which allow us to select the scattering conditions for
maximum magnetic contrast~\cite{radu:jmmm:2006}. In this way  we
measured element specific hysteresis loops  as a function of
temperature which yield the H$_c$ and H$_{EB}$ shown in
Fig.~\ref{fig5}d and Fig.~\ref{fig5}e, respectively. Flipping the
circular helicity of X-rays as well as the magnetic field  allows
us to separate rotatable and frozen-in AF spins~\cite{rz:2008}
which are depth and laterally uncompensated (Fig.~\ref{fig5}c). By
contrast an ideal uncompensated monolayer assumed by the
Meiklejohn and Bean (M\&B) model~\cite{bean:1957} is essentially
depth and laterally compensated (see previous section)  for our
large AF layer, therefore it will not contribute to a  shift of
the macroscopic hysteresis loop.
Sensitivity to  monoatomic uncompensated M\&B  spins appears to be
provided by a more localized probe like x-ray magnetic circular
dichroism as debated in Refs.~\cite{ohldag:2001,tsunoda:2006}.

Measuring the reflected intensity for circularly right
($I^{\sigma+}$) and left ($I^{\sigma-}$) polarized x-rays while
sweeping the external field, one obtains a hysteresis loop
provided by the asymmetry ratio as a function of an external
field: $A (H)\equiv A=
(I^{\sigma+}-I^{\sigma-})/(I^{\sigma+}+I^{\sigma-})$. This
asymmetry ratio resolves vertical shifts of the hysteresis loops,
with respect to the magnetization axis. Now, the frozen-in and
rotatable AF uncompensated components  are extracted from the
positive ($A_{sat}^{pos}$) and negative ($A_{sat}^{neg}$) magnetic
saturation  values of asymmetry as:
$A^F=(A_{sat}^{pos}+A_{sat}^{neg})/2$ and
$A^R=(A_{sat}^{pos}-A_{sat}^{neg})/2$, respectively. These two
experimental observables are plotted in Fig.~\ref{fig5}c as a
function of temperature. The sum of rotatable and frozen-in spins
is seemingly constant (not shown) and extends through the N\'{e}el
temperature~\cite{radu:jmmm:2006,roy:2007,abrudan:2008}. Note that
the rotatable asymmetry component is much larger than the frozen
asymmetry. At the blocking temperature, however, 3\% of these
rotatable AF spins become frozen with a sharp characteristic
onset. Note that the absolute asymmetry values for the frozen-in
spins (Fig.~\ref{fig5}c)) is below 0.0015, which reflects an
exceptionally  high precision (not achieved before) for these
measurements. At lower temperatures a linear increase of the
frozen-in asymmetry is clearly observed and correlates with the
in-plane AF coherence length. The direct relation between the size
of AF domains and the uncompensated spins is intrinsic to the
Malozemoff
model~\cite{malozemoff:1987,malozemoff:1988:1,malozemoff:1988:2}.
Calculating the H$_{EB}$ for this system within the Malozemoff
model~\cite{rz:2008} one would expect it to be about 600~Oe.
 The  measured H$_{EB}$ value of about 20~Oe is still
30 times lower.

 H$_{c}$ increases linearly as a function of temperature,
confirming a strong interfacial coupling. The exchange bias shows,
however, a very different behavior. After a sharp onset at the
blocking temperature (T$_B$=280~K), it increases linearly towards
low temperatures. The blocking temperature can be correlated to
the temperature where the AF domains achieve stability against the
exchange interaction with the FM layer as observed by neutron
scattering (Fig.~\ref{fig5}b). This origin of the blocking
temperature can be inferred from the  M\&B model. There, the
blocking temperature is always lower than the N{\'e}el
temperature, and is essentially governed by the magnitude of the
AF anisotropy energy and interfacial exchange energy. The
temperature where the AF (effective) anisotropy become strong
enough to resist the rotation (remagnetization) of the
ferromagnetic spins, is defined as blocking
temperature~\cite{bean:1957,rz:2008}.

The astounding  correlations between the temperature evolution of
the AF domain size, frozen-in spins, and value of exchange bias
are shown in Fig.~\ref{fig5}. The temperature dependence of the
H$_{EB}$ and of frozen-in spins is correlated with  the average AF
domain size and orientation. Characteristic features of three
different models can be inferred from these data: the origin of
the blocking temperature can be described  by the M\&B model, the
formation of AF domains are intrinsic to the DS and Malozemoff
models, and the linear dependence between the AF frozen spins and
the AF domain sizes are characteristic to the Malozemoff model,
demonstrating their limitations. These features, including the
noncollinearity between the AF and FM spins, can  all be accounted
for by the Spin Glass (SG)
model~\cite{radu:jpcm:2006,rz:2008}. 

\section{Conclusions}

In conclusion, we have investigated an archetypal exchange bias
bilayer by using complementary neutron and x-ray diffraction
techniques. An almost anisotropic  orientation of  AF domains is
observed, thus, clarifying the origin of the reduction of the
exchange bias field in epitaxially grown CoO/FM bilayers by
several orders of magnitude.   The blocking temperature for the
exchange bias is the temperature where the antiferromagnetic
domains achieve stability against the exchange interaction with
the FM layer. By contrast, at the  N\'{e}el temperature the AF
system develops a long range order against thermal fluctuations.
At low temperatures, the antiferromagnetic domains are not stable
upon magnetization reversal, which is directly identified as one
contribution to the training effect. Uncompensated frozen-in spins
are found to be remarkably well correlated with the
antiferromagnetic domains and the exchange bias field. This
strongly supports a mechanism for exchange bias caused by
interfacial uncompensated spins.




\end{document}